\documentclass[12pt]{iopart}

\begin{document}
\newcommand{\Od}{{\cal O}}
\def\lsim{\raisebox{-.4ex}{$\stackrel{<}{\scriptstyle \sim}$\,}}
\def\gsim{\raisebox{-.4ex}{$\stackrel{>}{\scriptstyle \sim}$\,}}

\input epsf 
\title{Some model-independent
phenomenological consequences of flexible brane worlds}

\author{J.A.R. Cembranos}

\address{Department of Physics and Astronomy, University of California, Irvine,
CA 92697 USA}
\author{A. Dobado and A.L. Maroto}

\address{Departamento de F\'{\i}sica Te\'orica, Universidad Complutense de
Madrid, 28040, Madrid, Spain}

\begin{abstract}
In this work we will review the main properties of brane-world models 
with low tension. Starting from very general principles, it is possible to 
obtain an effective action for the relevant degrees of freedom at low
energies (branons). 
Using the cross sections for high-energy processes involving branons,  
we set bounds on the different 
parameters appearing in these models. We also  show that branons
provide a WIMP candidate for dark matter in a natural way. We consider
cosmological constraints on its thermal and non-thermal relic abundances. 
We derive direct detection limits and compare those limits with
 the preferred parameter region in the case in which the EGRET 
excess in the diffuse galactic gamma rays is due to dark matter annihilation. 
Finally we will discuss 
the constraints coming from the precision tests of the Standard Model 
and the muon anomalous 
magnetic moment.

\end{abstract}


\section{Introduction}
In the brane-world scenario \cite{ADD} our universe is
understood as a three-brane embedded in a higher dimensional space.
If the extra space is compact, the large radius $R$ of the $n$ extra dimensions
implies that the fundamental gravity scale in $D=4+n$ dimensions $M_D$
can be much lower than the Planck scale $M_P$. Indeed $M_P^2=M_D^{2+n}R^n$.
Thus typically it is possible to have weak-scale gravity $M_D \simeq 1$ TeV for
$n=2$ and $R\sim 1$ mm or $n=7$ and $R\sim 10$ Fermi. 
However, apart from $M_D$, there is another fundamental scale in these 
models which is given by the finite tension of the brane $\tau=f^4$. 
The relative size of these two scales determines two completely different 
 regimes. In the rigid-brane limit ($f\gg M_D$), 
the new relevant degrees of freedom are the Kaluza-Klein (KK) modes of fields
propagating in the extra dimensions. In the flexible-brane limit ($f\ll M_D$),
the KK modes decouple from the Standard Model (SM) fields \cite{GB,rad} and the only
new relevant degrees of freedom are those corresponding to the brane
fluctuations (branons). In this work we will review the main phenomenological
consequences of brane-world models in this latter case.
 
\section{Branons versus Kaluza-Klein gravitons}
Let us consider  our four-dimensional space-time 
$M_4$ to be
embedded in a $D$-dimensional bulk space whose coordinates 
will be denoted by $(x^{\mu},y^m)$, where 
$x^\mu$, with $\mu=0,1,2,3$, correspond to the 
ordinary four dimensional space-time and $y^m$, with 
$m=4,5,\dots,D-1$, are coordinates of the compact extra
space.

From the point of view of particle physics, the first new effects of  
brane-worlds are related to the KK mode expansion of the bulk gravitational field:
\begin{equation}
g_{\mu\nu}(x,\vec y)=\sum_{\vec{k}}
g_{\mu\nu}^{\vec k}(x)
e^{i  \vec k . \vec y/ R} 
\label{KKmode} \end{equation}
where a toroidal compactification has been assumed for simplicity and
 $\vec k$ is a $n$
dimensional vector with components $k^m=0,1,2,...$. 
Therefore a bulk graviton
can be understood as a KK tower of four dimensional massive gravitons 
with
masses of the order of $k/R$ with $k$ being any natural number (for the
$n=1$ case) so that the distance in the mass spectrum 
between two consecutive KK gravitons is 
of the order of $1/R$. This means in particular that the KK graviton 
spectrum
can be considered as approximately continuous for large extra 
dimensions. In
principle we expect two kinds of effects from the KK graviton tower, namely
graviton production and virtual effects on other particle production or
observables (see for instance \cite{HS} and references therein). The rates for
the different processes can be computed by linearizing the bulk gravitational
field and by coupling the graviton field to the SM energy-momentum tensor
$T_{SM}^{\mu\nu}$. Then expanding the gravitational field in terms of the KK
modes one finds the corresponding Feynman rules. 

However, apart from the existence of KK modes, the fact that the brane
is a physical object with finite tension requires the existence
of new fields which parametrize the brane fluctuations \cite{DoMa}. 
For simplicity we will assume that the bulk 
metric tensor takes the following form:
\begin{eqnarray}
ds^2=\tilde g_{\mu\nu}(x)W(y)dx^\mu dx^\nu- g'_{mn}(y)dy^m dy^n
\label{metric}
\end{eqnarray}
where the warp factor is normalized as $W(0)=1$.

Working in the probe-brane approximation, our
3-brane universe is moving in the background metric given 
by (\ref{metric})
which is not perturbed by its presence. 
The position of the brane in the bulk can be parametrized as
$Y^M=(x^\mu, Y^m(x))$, and  we assume for simplicity that the ground
state of the brane corresponds to $Y^m(x)=0$. 

In the simplest case in which the metric is not
warped along the extra dimensions, 
i.e. $W(y)=1$,  
the transverse brane fluctuations are massless and
they can be parametrized by the Goldstone bosons fields 
$\pi^\alpha(x),\; \alpha=4,5, \dots D-1$  (branons).  
 In that case we can choose the $y$
coordinates   so that the branon fields are
proportional to the extra-space coordinates:
$\pi^\alpha(x)
=f^2\delta_m^\alpha Y^m(x)$ ,
where the proportionality constant is related to the brane 
tension $\tau=f^4$ \cite{DoMa}.

In the general case, the curvature generated by the
warp factor explicitly breaks the traslational 
invariance in the extra space. Therefore branons  acquire a mass 
matrix which is given precisely by the bulk Riemann tensor
evaluated at the brane position:
\begin{eqnarray}
M^2_{\alpha\beta}=\tilde g^{\mu\nu}R_{\mu\alpha\nu\beta}\vert_{y=0}
\end{eqnarray}

The fact that the brane can fluctuate implies that the actual
metric on the brane is no longer given by 
$\tilde g_{\mu\nu}$, but by the induced metric which includes
the effect of warping through the mass matrix:
\begin{eqnarray}
g_{\mu\nu}(x,\pi)
=
\tilde g_{\mu\nu}(x)\left(1+\frac{M^2_{\alpha\beta}\pi^\alpha \pi^\beta}{4f^4}\right)
-\frac{1}{f^4}\partial_{\mu}\pi^\alpha
\partial_{\nu}\pi^\alpha +\Od(\pi^4), 
\end{eqnarray}
The dynamics of branons can be obtained from the 
 Nambu-Goto action by introducing the above expansion.
In addition, it is also possible to get their couplings to
the ordinary particles just by replacing the space-time by
the induced metric in the Standard Model action.
Thus we get, up to quadratic terms in the branon fields: 
\begin{eqnarray}
S_{Br} 
&=&\int_{M_4}d^4x\sqrt{\tilde g}\left[\frac{1}{2}
\left(\tilde g^{\mu\nu}\partial_{\mu}\pi^\alpha
\partial_{\nu}\pi^\alpha
-M^2_{\alpha\beta}\pi^\alpha \pi^\beta
\right)\right.\nonumber \\
&+&\left.\frac{1}{8f^4}\left(4\partial_{\mu}\pi^\alpha
\partial_{\nu}\pi^\alpha-M^2_{\alpha\beta}\pi^\alpha \pi^\beta 
\tilde g_{\mu\nu}\right)
T^{\mu\nu}_{SM}\right]
\label{lag}
\end{eqnarray}

We can see that branons  interact with the  SM
 particles through their energy-momentum tensor.
The couplings are controlled by the brane 
tension scale $f$ and they are  universal very much like
those of gravitons. For large $f$, branons are therefore 
weakly interacting particles.

\begin{figure}[h]
\centerline{\epsfxsize=7.5cm\epsfbox{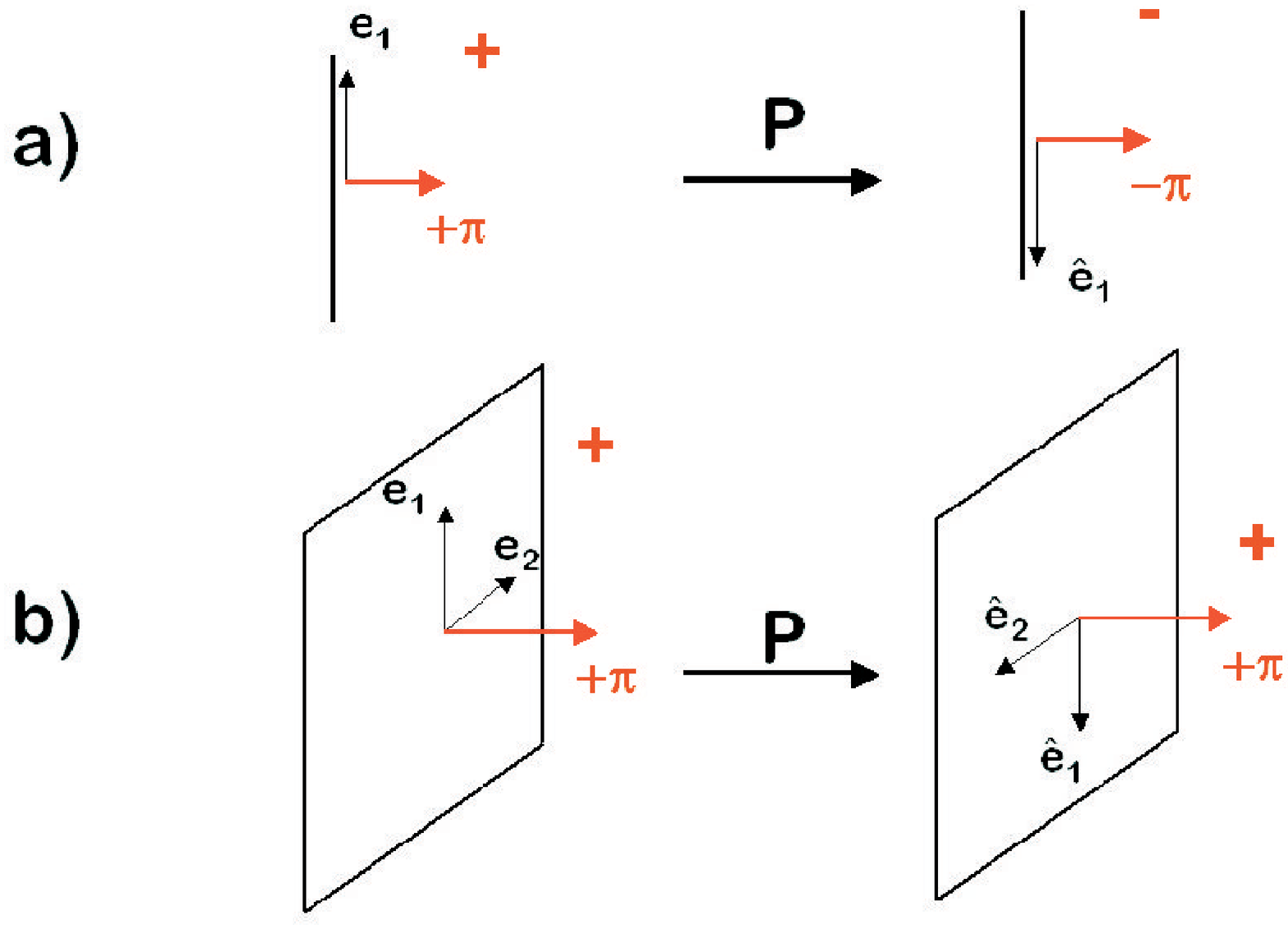}}
\vspace{.3cm}
\footnotesize {\bf Figure 1:}  Parity transformation on the brane for: a)  Odd-dimensional
brane. Branon fields change sign (pseudoscalars). b) Even dimensional brane (scalar
branons)
\end{figure}

The sign of the branon fields is determined by the orientation
of the brane submanifold in the bulk space. Under a parity 
transformation on the brane ($x^i\rightarrow -x^i$), the orientation
of the brane changes sign provided the ordinary space has an odd number
of dimensions, whereas it remains unchanged for even spatial dimensions
(see Fig.1). In the case in which we are interested with three ordinary spatial
dimensions, branons are therefore pseudoscalar particles.
If the gravitational sector of the theory respects parity on the brane,  
then  branons always couple 
to SM particles by pairs, which ensures that they are 
stable particles. This fact can have important consequences
in cosmology as we show below.

\section{Phenomenology in colliders}

The main  processes in colliders in which branons could contribute in a
relevant way are the single photon and single-Z production in $e^+e^-$
colliders (LEP) and the monojet and single-photon production in hadron colliders 
such as Tevatron \cite{L3}, \cite{ACDM}. 
In those processes, a pair of branons are produced in the
final state which are not detected and therefore they appear as 
missing energy and momentum. 
We have computed the limits on $(f,M)$ imposed by the 
absence of any deviation from the SM predictions in such processes. The results
are summarized in Fig. 2. 
\begin{figure}[h]
\centerline{\epsfxsize=4.5cm\epsfbox{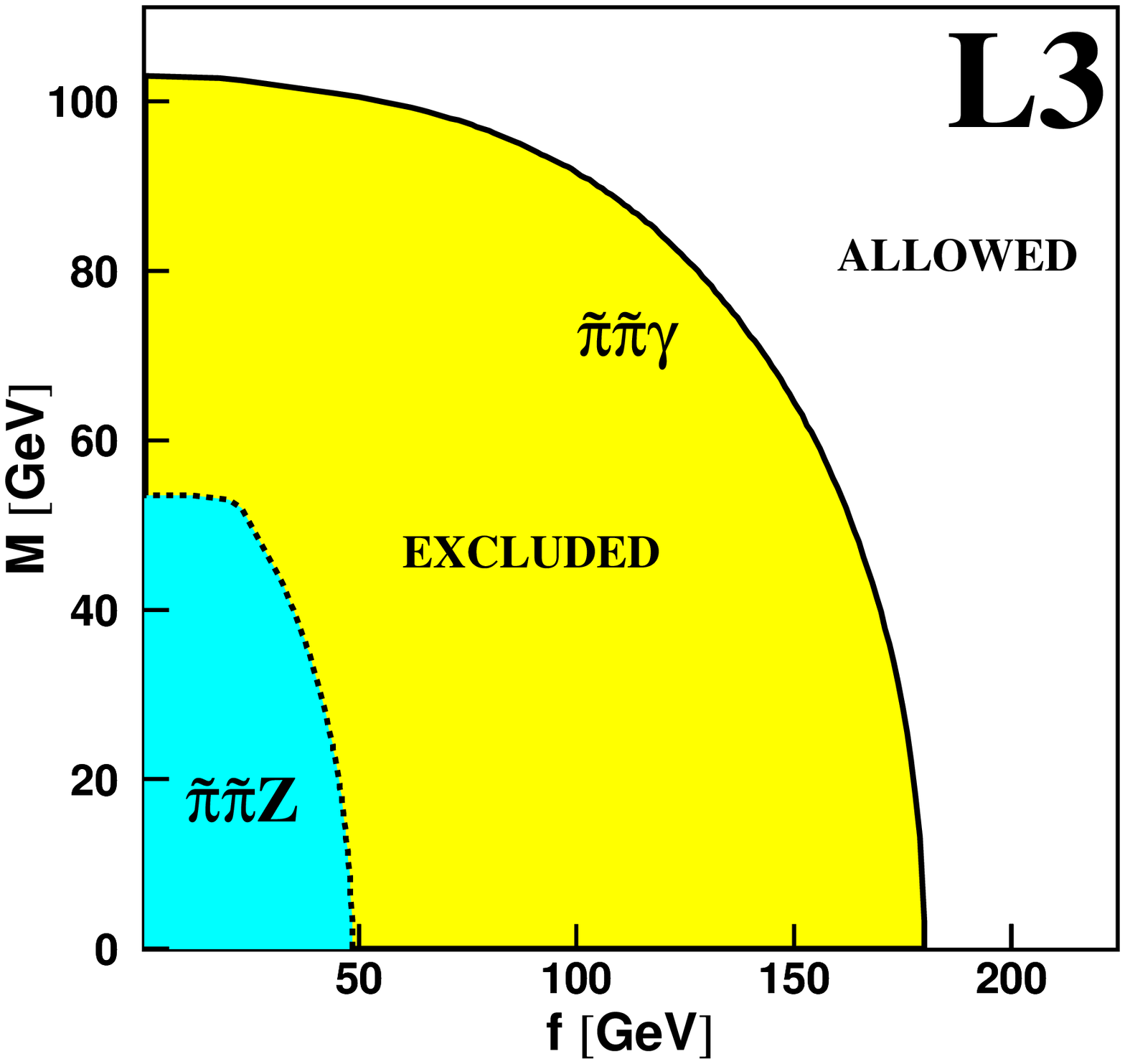}
\epsfxsize=6cm\epsfbox{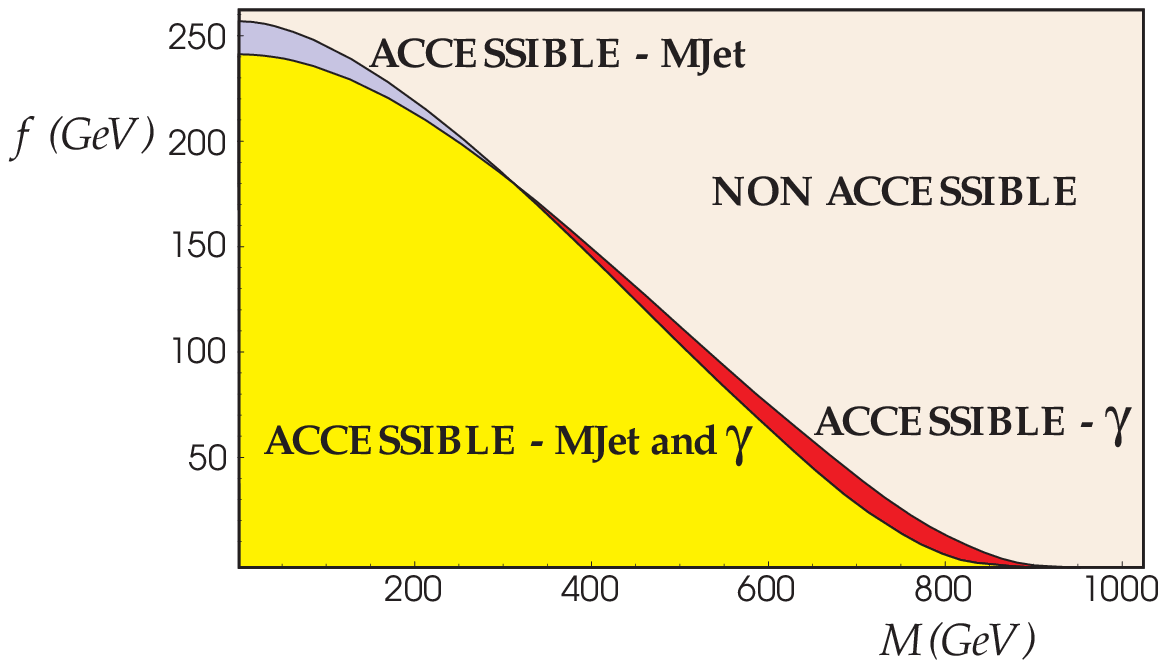}
}
\vspace{.3cm}
\footnotesize {\bf Figure 2:} 
Collider limits on branon parameters from single-photon and
single-Z processes at LEP (L3)  \cite{L3} (left). Limits from monojet 
and single-photon processes at Tevatron-I \cite{ACDM} (right)
\end{figure}

We have also estimated the prospects for future hadron colliders (Tevatron-II and
LHC) and $e^+e^-$ colliders (ILC). In Fig. 3 we show the accessible regions
in the $(f,M)$ plane. 
\begin{figure}[h]
\centerline{\epsfxsize=12.5cm\epsfbox{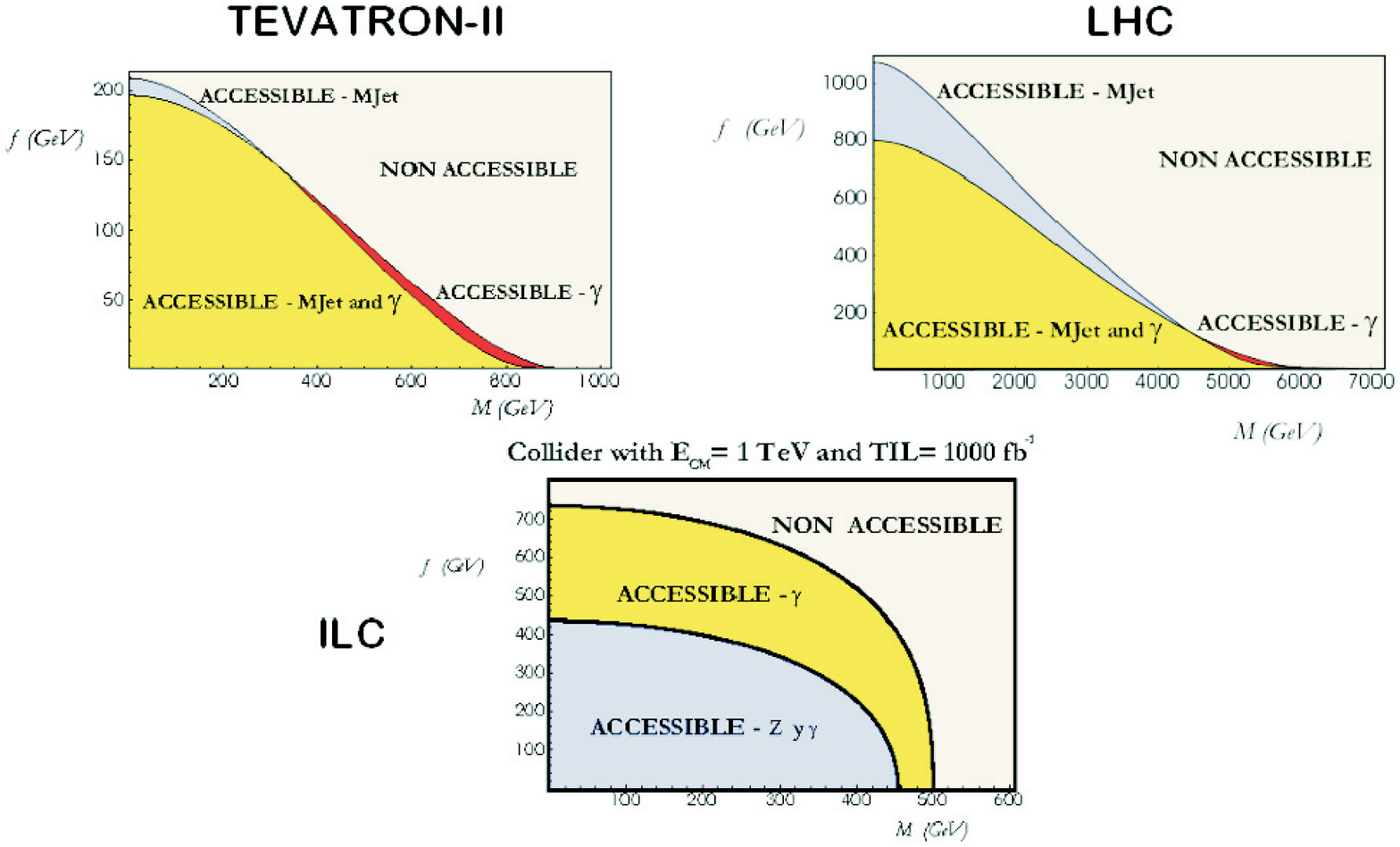}
}
\vspace{.3cm}
\footnotesize {\bf Figure 3:} 
Collider limits on branon parameters from monojet and single-photon 
processes for Tevatron-II and LHC and single-photon and
single-Z processes for the International Linear Collider (ILC)
with typical values for the center of mass energy of 1 TeV and total integrated luminosity of 
1000 fb$^{-1}$.  
\end{figure}

\section{Branons as dark matter}

As commented before branons are stable and weakly interacting particles, 
this makes them natural dark matter candidates \cite{CDM}.
\begin{figure}[h]
\centerline{\epsfxsize=10cm\epsfbox{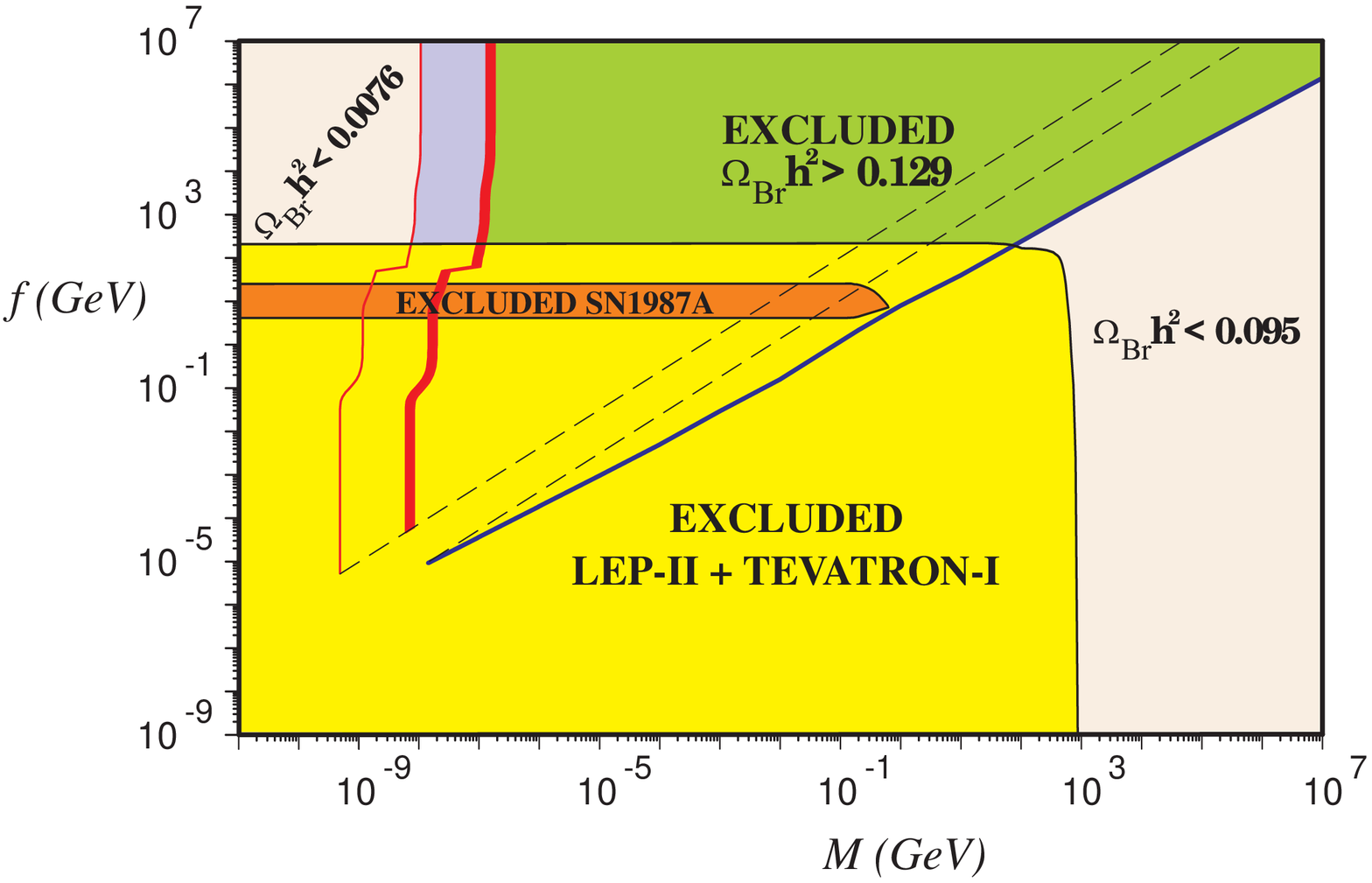}} 
{\footnotesize {\bf Figure 4:} 
Thermal relic abundance in the $f-M$ plane for a model with one branon of
mass $M$. The two lines on the left correspond to the 
 WMAP first year measurements: 
$\Omega_{Br}h^2=0.0076$ and $\Omega_{Br}h^2=0.129 - 0.095$ curves
for hot-warm relics, whereas the right line corresponds to the
latter limits for cold relics (see \cite{CDM} for details). The
lower area is excluded by single-photon processes at LEP-II
\cite{ACDM} together with monojet signal at Tevatron-I
\cite{ACDM}. The astrophysical constraints are less restrictive
and they mainly come from supernova cooling by branon emission
\cite{CDM}.} 
\end{figure}

 In \cite{CDM} the thermal relic branon abundance has been calculated
in two cases: either
relativistic branons at freeze-out (hot-warm) or non-relativistic
(cold), and assuming that the evolution of the universe is
standard for $T<f$ (see Fig. 4 for comparison with the WMAP first year limits). 

However, branons can also act as non-thermal dark 
matter very much in the same way as axions. Indeed, if the maximum
temperature reached in the universe is smaller than the branon
freeze-out temperature $T_f$, but larger than the explicit symmetry
breaking scale $(Mf^2R)^{1/2}$, then branons can be considered as 
essentially massless
particles decoupled from the rest of matter and radiation. In such
a case, there is no reason to expect that  the brane initial position
$Y_0$ should coincide with the potential minimum, but in general
we would have $Y_0\simeq R$. Thus as the universe cools down, the 
 brane can start oscillating coherently around the equilibrium position,
in a completely analogous way to the axion misalignment mechanism.
The energy density of those oscillations scales as that of  non-relativistic
matter, and can be estimated as \cite{NT}:
\begin{eqnarray}
\Omega_{Br}h^2\simeq  \frac{6.5\cdot 10^{-20}N}
{\mbox{GeV}^{5/2}} f^4\,R^2\,M^{1/2}
\end{eqnarray}
Limits on the different parameters in such
a case can be found in Fig 5. 

\begin{figure}[h]
\centerline{\epsfxsize=8cm\epsfbox{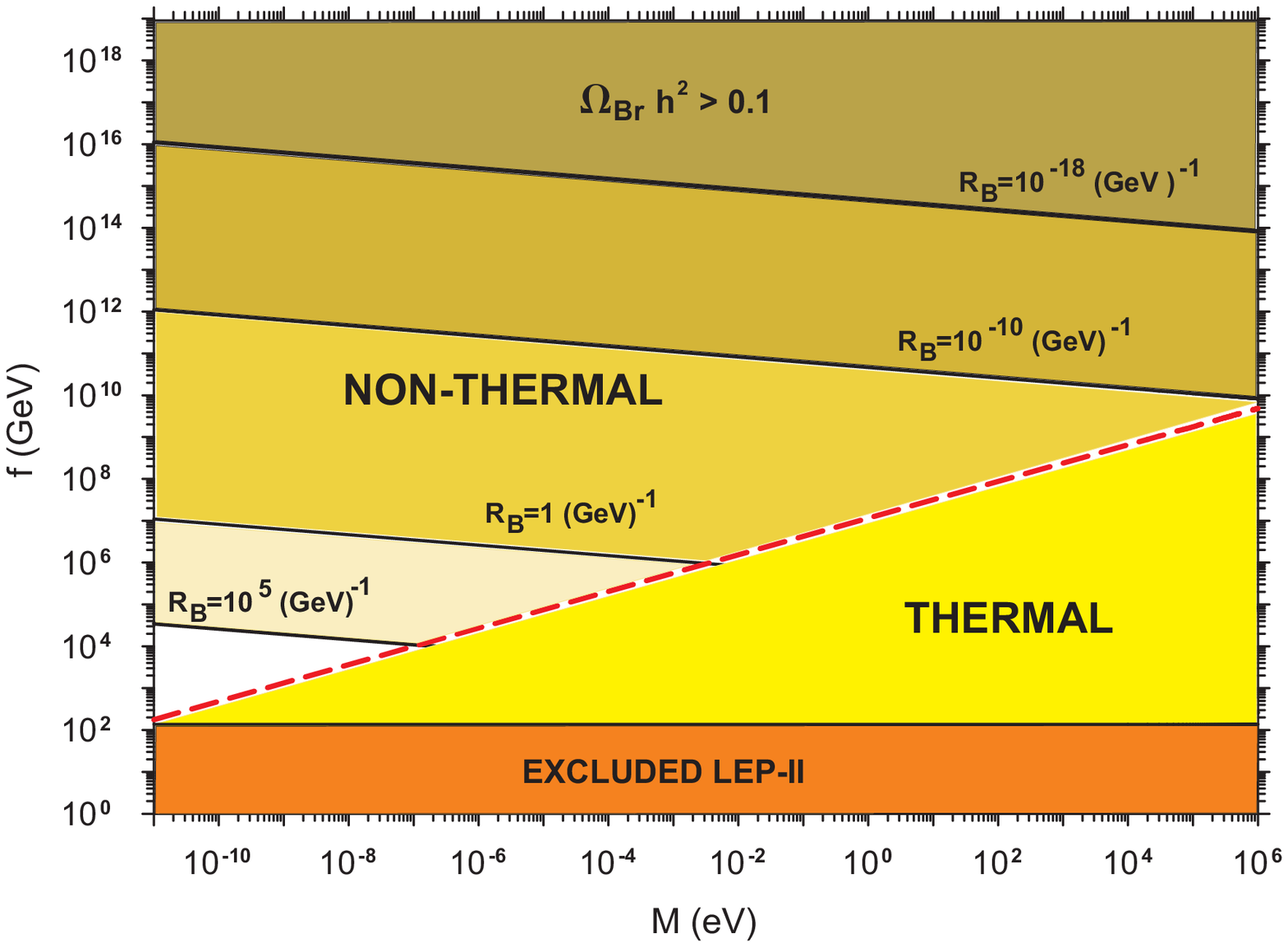}} 
\vspace{.3cm}
\footnotesize {\bf Figure 5:} Thermal vs. non-thermal branons regions in the $f-M$ plane. 
The dashed (red) line separates the 
two regions and corresponds to  $T_f\simeq (MM_P)^{1/2}$. The continuous (black) lines correspond to
$\Omega_{Br}h^2\simeq 0.1$ for different values of the radius of the extra dimensions
$R=10^5,\, 1,\,10^{-10},\, 10^{-18}$ GeV$^{-1}$.
\end{figure}

If branons constitute the dark matter of the universe, they will make up the 
galactic halo and could also be  detected by direct or indirect detection experiments.
In the direct case, branons interact with the nuclei in the detector and
the nucleus recoil energy can be measured. Several underground experiments
set limits on the branon mass and the elastic branon-nucleon cross section
at zero momentum transfer $\sigma_n$ \cite{CDM} (see Fig. 6).

\begin{figure}[ht]
\centerline{\epsfxsize=9cm\epsfbox{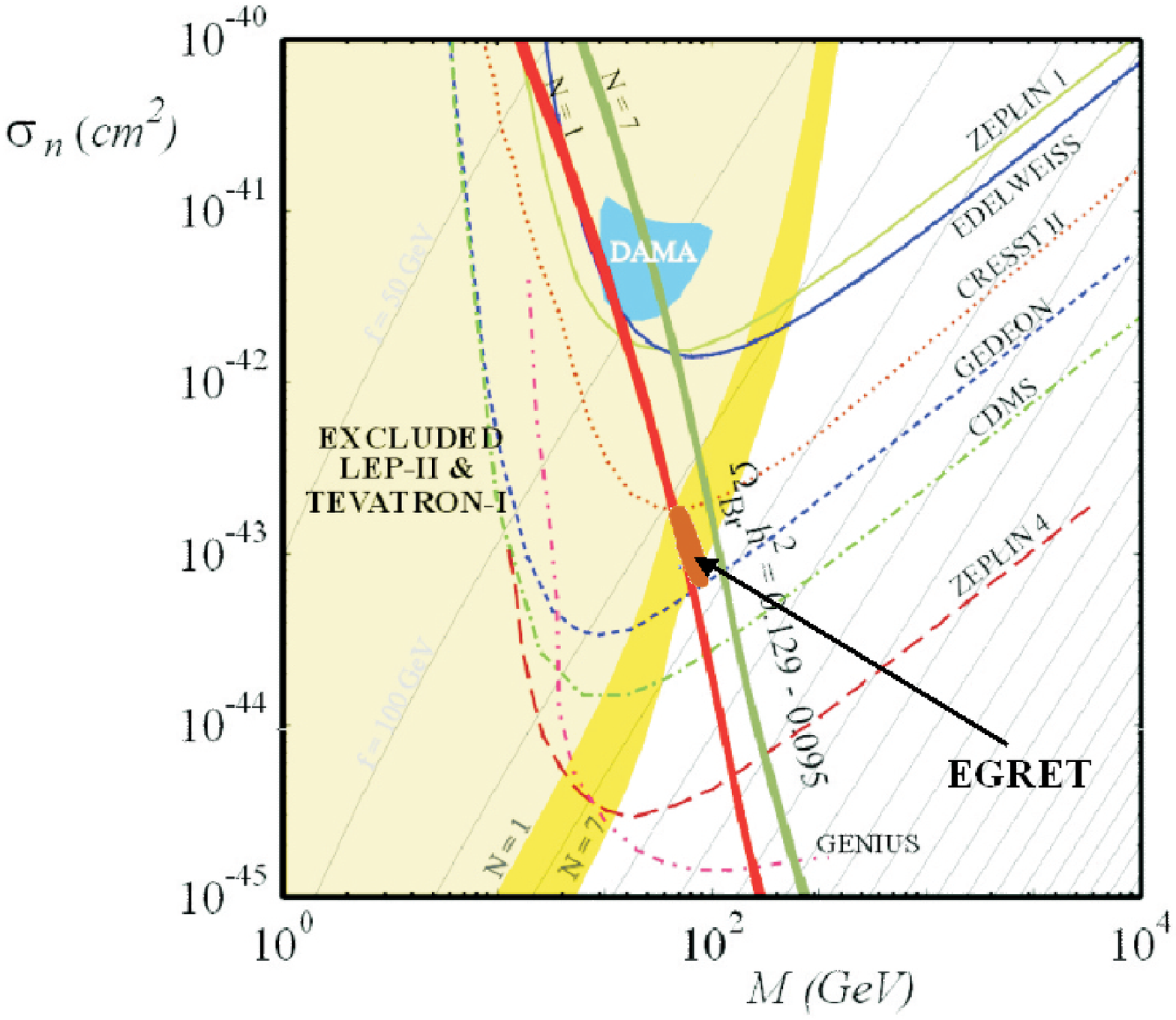}} 
{\footnotesize {\bf Figure 6:} Elastic branon-nucleon cross section 
$\sigma_n$ in terms
of the branon mass. The thick (red and green) line corresponds to the
$\Omega_{Br}h^2=0.129-0.095$ curve for cold branons in Fig. 2 
from $N=1$ and $N=7$ branon species. The shaded areas are the LEP-II
and Tevatron-I exclusion regions. The
solid lines correspond to the current limits on the
spin-independent cross section from direct detection experiments.
The discontinuous lines are the projected limits for future
experiments. Limits obtained from \cite{dmtools}.
The EGRET region correspond to parameter space for which dark matter 
annihilation in the galactic halo could account for the 
gamma-ray excess observed by EGRET \cite{Boer}.}
\end{figure}
Branons could also be detected indirectly: their annihilations in
the galactic halo can give rise to pairs of photons or $e^+ e^-$
which could be detected by $\gamma$-ray telescopes such as MAGIC
or GLAST or antimatter detectors (see \cite{AMS} for an estimation
of positron and photon fluxes from branon annihilation in AMS).
Annihilation of branons trapped in the center of the sun or the
earth can give rise to high-energy neutrinos which could be
detectable by high-energy neutrino telescopes. 

Dark matter annihilation  has been proposed as
a possible explanation for the excess of flux above 1 GeV 
in the EGRET data of the diffuse galactic gamma rays. 
For that purpose a WIMP mass around 50-70 GeV would be required.
In Fig. 6 we include the parameter region calculated from \cite{Boer}

\section{Virtual branon effects}
Branons can  also have phenomenological consequences through virtual 
effects \cite{rad}. Probably the most immediate effect of virtual branons is the
suppression of the coupling of SM particles and the KK modes of bulk
fields like the graviton. When branon fluctuations are taken into
account this effective coupling is described by the action:
\begin{eqnarray}
S_h&=& \frac{1}{\bar M_P} \sum_p  \int_{M_4}d^4x
h_{\mu\nu}^{(p)}(x) T^{\mu\nu}_{SM}(x)f_p(\pi)
\end{eqnarray}
where $\bar M_P^2 \equiv M_P^ 2/ 4 \pi$ is the squared reduced
Planck mass. The $f_p(\pi)$ couplings appear 
due to the fact that the brane is no more sitting at $\pi=0$, but
can fluctuate around this point. Now the branons fields can be
integrated out in the usual way to find:
\begin{eqnarray}
S_h&=& \frac{1}{\bar M_P} \sum_p  \int_{M_4}d^4x
h_{\mu\nu}^{(p)}(x) T^{\mu\nu}_{SM}(x)\langle f_p(\pi)\rangle
\end{eqnarray}
where the $f_p$ expectation value is given by:
\begin{eqnarray}
\langle f_p(\pi) \rangle = \int [d\pi] e^{i S_{eff}^{(2)}[\pi]}
f_p(\pi)
\end{eqnarray}

To compute the path integral above, we need to know the precise
form of the $f(\pi)$ functions which depends on the 
geometry of the extra dimensions. For the simplest case of torus compactification,
the effective action becomes:
\begin{eqnarray}
S_h&=& \frac{1}{\bar M_P} \sum_{\vec k}  \int_{M_4}d^4x g_{\vec
k}h_{\mu\nu}^{(\vec k)}(x) T^{\mu\nu}_{SM}(x)
\end{eqnarray}
In other words, the effect of branon quantum fluctuations amounts
to introducing the KK mode dependent couplings $g_{\vec k}$ which
are given for toroidal compactification by
\begin{equation}
g_{\vec k}=\exp  \left(-\frac{\Lambda^ 2}{32\pi^2R^2 f^4}
\sum_{m=1}^N (k^m)^ 2\left[1-\frac{M_m^ 2}{\Lambda^
2} \log \left(\frac{\Lambda^ 2}{M_m^
2}+1\right)\right]\right)
\end{equation}
where we have introduce and ultraviolet cutoff $\Lambda$ in order 
to regularize the divergent integrals. 
Thus the coupling of SM matter to higher KK modes is exponentially
suppressed.  This result was first obtained in \cite{GB} for
massless branons by using an argument based on normal ordering.
In any case this coupling suppression  has very interesting
consequences from the phenomenological point of view. It improves
the unitarity behavior of the cross section for producing
gravitons from SM particles and, in addition, it solves the
problem of the divergences appearing even at the tree level when
one considers the KK  graviton tower propagators for dimension
equal or larger than two. Moreover whenever we have $v\equiv R f^
2 \ll \Lambda$, KK gravitons decouple from the SM particles, so
that at low energies the only brane-world related particles that
must be taken into account are branons. 

In order to study the effect of virtual branons on the SM
particles, it is useful to introduce the SM effective action
$\Gamma_{SM}^{eff}[\Phi]$ obtained after integrating out the
branon fields:
\begin{eqnarray}
e^{i\Gamma_{SM}^{eff}[\Phi]}=\int [d\pi] e^{iS_{SM}[\Phi,\pi]}
\end{eqnarray}
This effective action is in general a non-local and divergent functional
of the Standard Model fields and in order to obtain
finite results we have used a cut-off ($\Lambda$) regularization procedure
as before. The new
local divergent terms are organized in powers of the fields so that, 
at the level of two-point functions, we have \cite{rad}:
\begin{eqnarray}
\Delta {\cal L}^{(1)}=C_1 T^{\mu}_{SM\mu}
\end{eqnarray}
where $C_1=-N\Lambda^4/(16(4\pi)^2f^4)$ with $N$ the number 
of branon species. This term is proportional to the trace
of the SM energy-momentum tensor and accordingly amounts to a 
renormalization of SM masses. To second order, we get:
\begin{eqnarray}
\Delta {\cal L}^{(2)}=W_1 \,T^{\mu\nu}_{SM}T_{\mu\nu}^{SM}+
W_2\,T^{\mu}_{SM\mu}T^{\nu}_{SM\nu}
\label{eff}
\end{eqnarray}
where $W_1=2W_2=N\Lambda^4/(96(4\pi)^2f^8)$.  These counterterms
introduce new interaction vertices which are not present in the SM
Lagrangian. 
The most relevant ones could
be the four-fermion interactions or the fermion pair annihilation
 into two gauge bosons. We have used the data
coming from HERA, Tevatron and LEP
 on this kind of processes in order to set
bounds on the parameter combination $f^2/(\Lambda N^{1/4})$. The
results are shown in Table 1, where it is also
possible to find the prospects for  the future colliders mentioned
above.

\begin{table}[bt]
\begin{center}
\begin{tabular}{||c|c c c||} \hline
Experiment          & $\sqrt s$ (TeV) & ${\cal L}$ (pb$^{-1}$) &
$f^2/(N^{1/4}\Lambda)$ (GeV) \\ \hline
HERA$^{\,c}$        & 0.3             &  117                   & 52                           \\
Tevatron-I$^{\,a,\,b}$   & 1.8        &  127                   & 69                           \\
LEP-II$^{\,a}$      & 0.2             &  700                   & 59                           \\
LEP-II$^{\,b}$      & 0.2             &  700                   &
75                           \\ \hline
Tevatron-II$^{\,a,\,b}$  & 2.0        & $2 \times 10^3$        & 83                           \\ 
ILC$^{\,b}$         & 0.5             & $5\times 10^5$         & 261                          \\
ILC$^{\,b}$         & 1.0             & $2\times 10^5$         & 421                          \\ 
LHC$^{\,b}$     & 14              & $10^5$                 & 383
\\ \hline
\end{tabular}
\end{center}
{\footnotesize {\bf Table 1:} 
Limits from virtual branon searches at colliders (results at the
$95\;\%$ c.l.) The indices $^{a,b,c}$ denote the two-photon,
$e^+e^-$ and $e^+p$ ($e^-p$) channels respectively. The first four
analysis have been performed with real data, whereas the final
four are estimations. The first two columns correspond to the 
center of mass energy and total luminosity, and the third one 
corresponds to the lower
bound on $f^2/(N^{1/4}\Lambda)$.}
\end{table}

As we have commented above, $\Lambda$ is the cutoff which limits
the validity of the effective description of branon and SM
dynamics. This new parameter appears when dealing with branon
radiative corrections since the Lagrangian in (\ref{lag}) is not
renormalizable. A one-loop calculation with the new effective
four-fermion vertices coming from (\ref{eff}),  is equivalent to a
two-loop computation with the Lagrangian in (\ref{lag}), and
allows us to obtain the contribution of branons to the muon anomalous
magnetic moment \cite{muon}:
\begin{equation}\label{gb}
\delta a_\mu \simeq \frac{5\, m_\mu^2}{114\,(4\pi)^4}
  \frac{N\Lambda^6}{f^8}.
\end{equation}
 We can observe that the correction has the
right  sign and that
 it is thus possible to
 improve the agreement with the  experimental
value. In fact,  the  difference between the
experimental and the SM prediction \cite{BNL,gm2} measured 
at the $2.7\sigma$ level by the
Brookhaven $g-2$ collaboration is $\delta a_\mu \equiv a_\mu
({\rm exp}) - a_\mu ({\rm SM}) =(23.4 \pm 9.1)\times 10^{-10}$.
 Thus, we can estimate the preferred parameter region for
branon physics as:
\begin{equation}\label{g-2}
    6.0\; \mbox{GeV}\;\gsim\,\frac{f^4}{ N^{1/2}\Lambda^3}\,
\gsim\;2.2\; \mbox{GeV}\; (\,95\; \%\; c.l.\,)
\end{equation}
Another limitation to the branon parameters could be obtained from
electroweak precision measurements, which use to be very useful to
constrain models of new physics. The so called oblique corrections (the
ones corresponding to the $W$, $Z$ and $\gamma$ two-point
functions) use to be described in terms of the $S,T,U$ 
or the $\epsilon_1,\epsilon_2$ and $\epsilon_3$ parameters. 
The experimental values obtained by LEP
 are consistent with the SM prediction
for a light Higgs $m_H\leq 237$ GeV at 95 \% c.l. In principle, it
is necessary to know this parameter in order to put constraints on
new physics, but one can calculate the disfavored regions of parameters in
order to avoid fine tunings. We can estimate this area by
performing a computation of the parameter
$\bar\epsilon\equiv \delta M_W^2/M_W^2 - \delta M_Z^2/M_Z^2$,
 in a similar way as it was done for the first order correction coming from the
Kaluza-Klein gravitons in the ADD models for rigid branes
\cite{CPRS}. The experimental value of $\bar\epsilon$ obtained
from LEP is $\bar\epsilon=(1.27 \pm
0.16)\times 10^{-2}$. The theoretical uncertainties are one order
of magnitude smaller  and therefore, we can estimate
the constraints for the branon contribution at 95 \% c.l. as
$|\delta\bar\epsilon|\,\lsim\, 3.2\times 10^{-3}$ with the result
\cite{rad}:
\begin{equation}\label{eb}
    \frac{f^4}{ N^{1/2}\Lambda^3}\;\gsim\;3.1\;
\mbox{GeV}\; (\,95\; \%\; c.l.\,)
\end{equation}

\section{Conclusions}
In this work we have reviewed the main phenomenological consequences
of flexible brane worlds. The new effects can be described in a 
model-independent way by means of the low-energy effective action
for branons which depends essentially on only two parameters: the 
brane tension scale $f$ and the branon mass $M$. We have computed the  
limits on those parameters from LEP and Tevatron I data, and the
prospects for detection in future colliders. 

We have shown that branons are natural candidates for dark matter both
as thermal and non-thermal relics in different regions of the 
parameter space. We have obtained the cosmological limits from 
WMAP observations and also studied the prospects for detection of
branons  in the galactic halo from direct or indirect 
searches experiments. 

Finally, we have considered additional constraints on branon physics from
virtual effects, and those coming from electroweak precision observables.
The new two-loop diagrams involving branons are shown to give a non-vanishing
contribution to the muon anomalous magnetic moment, which could explain the
observed difference between the SM prediction and the measurements of the 
Brookhaven $g-2$ collaboration. 

\vspace{.5cm}

{\em Acknowledgments:} 
This work has been partially supported by DGICYT (Spain) under project numbers
FPA 2004-02602 and FPA 2005-02327, by  NSF grant No. PHY-0239817 and
Fulbright-MEC award.

\section*{References}

\end{document}